\documentstyle[multicol,aps,prb,floats,eqsecnum]{revtex}
\newcommand{\calH}{{\cal H}}
\newcommand{\calD}{{\cal D}}
\newcommand{\calO}{{\cal O}}
\newcommand{\psibar}{{\overline{\psi}}}
\newcommand{\Psibar}{{\overline{\Psi}}}
\newcommand{\gtilde}{{\tilde g}}

\newcommand{\zbar}{{\bar z}}
\newcommand{\wbar}{{\bar w}}

\newcommand{\li}{{\mbox{li}}}

\begin{document}
\draft                                        
\preprint{}                                   
\title{ Interaction effects on random Dirac fermions} 
\author{ Takahiro Fukui\cite{Email}}
 \address{Theoretical Physics, University of Oxford, 
          1 Keble Road, Oxford OX1 3NP, UK}
\date{\today}
\maketitle
\begin{abstract}
We study a Dirac fermion model with three kinds of 
disorder as well as a marginal interaction 
which forms the critical line of $c=1$ conformal field theory.
Computing scaling equations by the use of
a perturbative renormalization group method,
we investigate how such an interaction affects
the universality classes 
of disordered systems with non-interacting fermions.
We show that 
some specific fixed points are stable against an interaction,
whereas others are unstable and 
flow to new random critical points with a finite interaction.
\end{abstract}


\begin{multicols}{2}
\section{Introduction}

Anderson localization has attracted renewed interest in 
recent years, especially due to the discovery 
by Altland and Zirnbauer\cite{AltZir} of novel universality 
classes for disordered systems.
In addition to the well-known Wigner-Dyson classification of 
the random matrices, 
seven new classes have been established. 
They can be found in quasiparticle systems described by
Bogoliubov-de Gennes Hamiltonians,  
or in systems with sublattice (chiral) symmetry, 
studied for the first time by Gade and Wegner.\cite{GadWeg}
As is well-known, symmetry properties play a crucial role in the 
localization problems. The key symmetry of the new classes 
is  a particle-hole symmetry, which makes the spectrum symmetric 
with respect to  the zero-energy.

The new universality classes have turned out to be quite useful in studying 
problems of quasiparticle localization in dirty $d$-wave 
superconductors.\cite{SFBN,BCSZ,ASZ} 
Random-bond models on bipartite lattices such as 
the random flux model,\cite{Fur} 
the random hopping model with $\pi$ flux per 
plaquette,\cite{Hat,Fuk,GLL,FabCas,RyuHat,MDH} 
graphite sheets,\cite{MDH} etc 
have also been examined successfully 
from the point of view of the chiral classes.
What is interesting is that many of these models 
can be described, near their critical points,
by Dirac fermions with appropriate symmetries.
Surprisingly, Dirac fermion models, 
with a slight extension to multi-species cases, 
exhaust all possible universality classes found so far, 
and detailed classification has been given 
by Bernard and LeClair.\cite{BerLeC}

On the other hand,
a disordered Dirac fermion model has been proposed as an effective model
for the integer quantum Hall transition.\cite{LFSG} 
Actually, the Chalker and Coddington network model\cite{ChaCod}, 
which is believed to describe the transition correctly, 
yields disordered Dirac Hamiltonian in a continuum limit.\cite{HoCha}
Some specific fixed points of the model, e.g.
the fixed line due to a random vector potential,\cite{LFSG,MCW,Ber} 
or the fixed point due to a random mass\cite{Ber,DotDot,Sha,SenFis,BSZ} 
have been analyzed in a weak-coupling regime.

Although the disordered Dirac fermions are one of the simplest models, 
many problems still remain to be explored.
One of them is to clarify strong coupling regime of the model.
It is quite necessary to find out a generic 
fixed point of the Dirac fermions with generic disorder, 
because it should be 
a field theory describing the integer quantum Hall transition.
Moreover, even the well-known fixed line formed by 
a random vector potential should be in a different phase, 
in a strong coupling regime, 
from the one expected by conventional weak-coupling
approaches.\cite{CMW,Cas}
However, recent developments enable us to 
discuss strong coupling aspects by the use of 
the renormalization group method or 
a variational method.\cite{CarDou,HorDou}  

Another interesting problem is to clarify effects of interaction 
on the localization properties. 
Studies on the random bond Ashkin-Teller model have
been reported by Dotsenko and Dotsenko.\cite{DotDotAT}
Describing the model by Majorana fermions and
computing one loop renormalization group (RG) equations,
they have found that the Harris criteria does not hold any longer 
in this model.
Their question is whether 
this feature is common in two-dimensional systems.
More generally, an interplay between interactions and randomness
has been extensively studied.\cite{FukYam}

The notion of the new universality classes, mentioned-above,
for the quasiparticle systems as well as random hopping models on bipartite
lattices are only for {\it non-interacting} systems, and
if models include interactions and randomness simultaneously, 
interactions in general break the symmetry 
which the models without interactions should have.
Therefore, it is quite interesting to study how interactions 
affect the symmetry properties of the newly discovered universality classes.

In this paper, we study a Dirac fermion model with three kinds of 
disorder as well as a marginal interaction which forms the 
critical line of the $c=1$ conformal field theory,
in order to clarify whether the universality 
classes for non-interacting theories are stable against such an interaction. 
Conversely, we might say that we study
how $c=1$ theory are changed by randomness.
We calculate one-loop RG equations by the use of the replica trick.
It turns out that there is an interesting interplay 
between the marginal interaction and randomness.
Namely, the universality classes for non-interacting fermions 
are stable again this interaction in some cases, but in other cases, 
the interaction survives and show the critical line like the $c=1$ theory. 

This paper is organized as follows.
In the next section, we define the model 
and derive the scaling equations. From Sec. III to Sec. VI, 
we study four kinds of specific fixed points in detail.
In Sec. VII, we summarize the results and 
briefly discuss a generic fixed point. 

\section{The model and scaling equations}

The model we will study in this paper is described by the 
following Dirac Hamiltonian in two dimensions 
with quenched disorder as well as an interaction,
\begin{eqnarray}
\calH&&=\psibar 
\left[ i\gamma_\mu \left(\partial_\mu-iA_\mu\right) +M\sigma_3+V
\right]\psi+\calH_{int}
\nonumber\\
&&=\psibar h\psi +\calH_{int} ,
\label{Ham}
\end{eqnarray}
where $M(x)$, $A_\mu(x)$, and $V(x)$ are a random mass, a random vector 
potential, and a random scalar potential, respectively, and
$\psi$ and $\psibar$ fields are two-component Dirac fields,
whose chiral components are defined by
\begin{eqnarray}
\psibar=-i\left(\psibar_L,\psibar_R\right), \quad
\psi=\left( \begin{array}{c}\psi_R\\\psi_L\end{array}\right) .
\end{eqnarray}
With respect to these notations 
the interaction $\calH_{int}$ is defined here as
\begin{eqnarray}
\calH_{int}=2gJ_RJ_L  ,
\label{Int}
\end{eqnarray}
where R- and L-currents are defined by
\begin{eqnarray}
J_R=\psibar_R\psi_R, \quad J_L=\psibar_L\psi_L .
\end{eqnarray}
Without disorder, this interaction is exactly marginal, and the model 
is on the critical line with continuously varying critical exponents 
of the $c=1$ conformal field theory. 

The model (\ref{Ham}) can describe critical behavior of various kinds of
lattice models.
One of well-known examples is 
the Ashkin-Teller model or equivalently the Baxter model.
Without randomness, it has been shown  \cite{Bax,KadBro,ShaAT} 
that the models form a critical line, and
in the scaling limit, they can be described by two kinds of Majorana
fermions or equivalently one Dirac fermion with the interaction (\ref{Int}).
The case $g=0$ corresponds to the decoupling point of two Ising spins,
giving just two independent Ising models.

If random bond interactions are introduced to this model, 
the Dirac fermion acquires a random mass term as in Eq. (\ref{Ham}).
Dotsenko and Dotsenko \cite{DotDot} have firstly studied the Ising model 
(the $g=0$ case) and shown that the random mass is marginally irrelevant, 
and it gives rise to the famous $\ln|\ln\tau|$ 
behavior of the specific heat,  where $\tau$ is the reduced temperature. 
In the case of the finite interaction, 
the same authors have shown \cite{DotDotAT} that
the positive interaction leads to the Ising fixed point $g=0$, 
whereas the negative interaction leads to  $1/(\ln|\ln\tau|)$ 
behavior of the specific heat.
This means that interaction and randomness couple together,
yielding new critical behavior.
 
Other lattice models are two-dimensional quantum systems 
of (nonrelativistic) fermion on bipartite lattices
such as a square lattice with $\pi$-flux per plaquette 
and a honeycomb lattice.
In such quantum systems, the interaction term (\ref{Int}) does not arise  
from interactions of the original lattice models, 
since one has to treat, in that case,  
a field theory of the Dirac fermion in three dimensions. 
One possibility is the coexistence of annealed randomness and quenched 
randomness: The former can serve as the interaction (\ref{Int}).
Actually, Belitz et. al. have stressed 
the importance of annealed randomness for systems
with quenched randomness and proposed a new mechanism for 
a metal-insulator transition.\cite{Kir}

The partition function is defined by
\begin{eqnarray}
Z=\int\calD\psi\calD\psibar\exp
\left[-\int d^2x\left(\calH-z\psibar\psi\right)\right], 
\label{ParFun}
\end{eqnarray}
where $z=E+i\epsilon$.
In order to compare the aspects of several fixed points, 
we will devote ourselves to the calculation of 
a two-point function and resultant 
density of state (DOS). For this purpose,
we introduced a source term $z\psibar\psi$ 
in the action functional (\ref{ParFun}).
Without the interaction (\ref{Int}), 
the two-point function is $\langle\psi\psibar\rangle=(h-z)^{-1}$,
where $h$ is defined in Eq. (\ref{Ham}).
In the present case, the interaction $\calH_{int}$ 
yields a self-energy $\Sigma$ and the two-point function can be expressed
as $\langle\psi\psibar\rangle=(h-\Sigma-z)^{-1}$.

Now we assume that the three kinds of randomness obey the Gaussian 
distribution with zero mean
\begin{eqnarray}
&&
\overline{A_\mu(x)A_\nu(y)}=g_A\delta_{\mu\nu}\delta(x-y), 
\nonumber\\ 
&&
\overline{M(x)M(y)}=g_M\delta(x-y),
\nonumber \\ 
&&
\overline{V(x)V(y)}=g_V\delta(x-y) .
\label{ProDis}
\end{eqnarray}
Then ensemble-average over disorder by the use of the replica trick
leads to the following Hamiltonian,
\begin{eqnarray}
\calH=&&\psibar_\alpha i\gamma_\mu\partial_\mu\psi_\alpha
-\frac{g_A}{2}\left(\psibar_\alpha\gamma_\mu\psi_\alpha\right)^2
-\frac{g_M}{2}\left(\psibar_\alpha\sigma_3\psi_\alpha\right)^2
\nonumber\\&& 
-\frac{g_V}{2}\left(\psibar_\alpha\psi_\alpha\right)^2
+2gJ_{R\alpha}J_{L\alpha}
-z\psibar_\alpha\psi_\alpha .
\label{RepHam}
\end{eqnarray}
where $\alpha=1,2,3...n$ denote the replicas, and
$J_{R(L)\alpha}=\psibar_{R(L)\alpha}\psi_{R(L)\alpha}$.

One can compute the 1-loop beta functions 
using the operator product expansion.
The unperturbed two-point functions read, 
\begin{eqnarray}
&&
\langle\psi_{R\alpha}(z)\psibar_{R\alpha'}(w)\rangle
=\frac{ \delta_{\alpha\alpha'} }{2\pi }\frac{1}{z-w},
\nonumber\\
&&
\langle\psi_{L\alpha}(\zbar)\psibar_{L\alpha'}(\wbar)\rangle
=\frac{\delta_{ \alpha\alpha'} }{2\pi }\frac{1}{\zbar-\wbar},
\label{TwoPoiFun}
\end{eqnarray}
and using these, we have the scaling equations
\begin{eqnarray}
&&
\frac{dg_A}{dl}
=\frac{1}{4\pi} \left( g_+^2-g_-^2 \right),
\nonumber\\
&&
\frac{dg_+}{dl}
=\frac{1}{\pi} \left[ 4g_A +(1-n)g_- +2g \right]g_+,
\nonumber\\
&&
\frac{dg_-}{dl}
=\frac{1}{2\pi} \left[ (2-n)g_+^2 -ng_-^2 +4gg_- \right],
\nonumber\\
&&
\frac{dg}{dl}=\frac{1}{\pi} gg_-,
\nonumber\\
&&
\frac{1}{z}\frac{dz}{dl}
=1+\frac{1}{2\pi} \left[ 2g_A +(1-n)g_+ -ng_- +2g \right],
\end{eqnarray}
where $g_\pm=g_V\pm g_M$, and 
$l=\ln L$ with the system length $L$.
Taking the replica limit $n\rightarrow0$, 
we reach the following RG equations 
which we will study in the following sections,
\begin{eqnarray}
&&
\frac{d\gtilde_A}{dl}=\gtilde_+^2-\gtilde_-^2,
\nonumber\\
&&
\frac{d\gtilde_+}{dl}=\left( \gtilde_A+\gtilde_-+\gtilde \right)\gtilde_+,
\nonumber\\
&&
\frac{d\gtilde_-}{dl}=  \gtilde_+^2+\gtilde\gtilde_- ,
\nonumber\\
&&
\frac{d\gtilde}{dl}= \gtilde\gtilde_-,
\nonumber\\
&&
\frac{1}{z}\frac{dz}{dl}=2-\gamma,
\label{RGeqs}
\end{eqnarray}
where we have denoted  the coupling constants as 
$\gtilde_A=4 g_A/\pi$, $\gtilde_\pm=g_\pm/\pi$, and
$\gtilde=2g/\pi$.  
When $g=0$, these equations are the same as those derived by Bernard.\cite{Ber}
The anomalous dimension $\gamma$ of $\psibar\psi$ is
\begin{eqnarray}
\gamma=1-
\left(\frac{\gtilde_A}{4}+\frac{\gtilde_+}{2}+\frac{\gtilde}{2}\right) .
\label{AnoDim}
\end{eqnarray}

The last equation in (\ref{RGeqs}) tells us 
that the energy $E=\mbox{Re}z$ grows according to
\begin{eqnarray}
\ln\frac{\Lambda}{E}=2(l-l_0)-\int_{l_0}^ldl\gamma,
\label{EneSca}
\end{eqnarray}  
where $E$ denotes the bare energy at the length scale $l_0=\ln L_0$
while $\Lambda$ is the renormalized energy 
at the scale $l=\ln L$ and serves as a cut-off. 
Since the scaling dimension of the field $\psibar\psi$ 
is given by (\ref{AnoDim}),
the ensemble-averaged DOS, which is the imaginary part of 
$\lim_{\epsilon\rightarrow0}\mbox{tr}\overline{\langle\psi\psibar\rangle}$, 
should obey the scaling law,
\begin{eqnarray}
\rho(L/L_0)\sim\exp\left( -\int^l_{l_0} dl \gamma\right).
\label{DosSca}
\end{eqnarray}
Ludwig et. al. have studied the same model without interaction.\cite{LFSG}
They examined  several specific fixed points of the model, although 
in the generic case with all possible randomness, scaling equations 
lead to a strong coupling regime, which is not accessible by 
a perturbative weak-coupling approach.
In the following sections, we will examine how several fixed points
known so far are affected by the interaction (\ref{Int}).

\section{The random vector potential}

In this section, we study the model
with the random vector potential and the interaction only.
First of all, it should be noted that without the interaction, 
the Hamiltonian has chiral U($n)\times$U($n$) symmetry, and therefore,
belongs to the class AIII.\cite{AltZir}
The interaction, however, breaks it up to $n$ copies of 
U($1)\times$U($1$) symmetry. 

It is known that the random vector potential is exactly marginal. 
On the other hand, the interaction is also exactly marginal.
It turns out that 
even if they exist at the same time, they remain to be exactly marginal.
Actually, by setting $\gtilde_+=\gtilde_-=0$, the scaling equations reduce to
\begin{eqnarray}
\frac{d\gtilde_A}{dl}=\frac{d\gtilde}{dl}=0 .
\end{eqnarray}
Therefore, the anomalous dimension is a constant given by
\begin{eqnarray}
\gamma
=1-\frac{\gtilde_{A0}}{4}-\frac{\gtilde_0}{2}
=1-\frac{g_{A0}+g_0}{\pi} ,
\end{eqnarray}
where $\gtilde_{A0}=4g_{A0}/\pi$ and $\gtilde_0=g_0/\pi$ are initial 
coupling constants at $l=l_0$.  
The energy flow is 
\begin{eqnarray}
\ln\frac{\Lambda}{E}=z(l-l_0),\quad z=1+\frac{g_{A0}+g_0}{\pi} ,
\end{eqnarray}
and hence the density of state is given by
\begin{eqnarray}
\rho(E)\sim\left(\frac{E}{\Lambda}\right)^{(2-z)/z} .
\end{eqnarray}
When $g_0=0$ this formula is just the same as that by Ludwig et. al.\cite{LFSG}
The interaction just modifies the exponent. 
However, from the point of view of symmetry properties, 
the model with finite interaction does not belong to AIII any longer.

\section{The random mass}

It is well-known that the random-bond Ising model can be described,
near the transition temperature, by the present Dirac model
with the random mass term only in Eq. (\ref{Ham}).\cite{DotDot,Sha} 
This model is invariant under O(2$n$) rotation
with respect to Majorana fermions, 
and therefore belongs to the class D.\cite{SenFis,BSZ}
The Ashkin-Teller model, which is composed of two Ising models
coupled with 4-spin interaction, 
\begin{eqnarray}
H=-\sum_{\langle i,j\rangle}
\left[
J_{ij}( \sigma_{1i}\sigma_{1j}+ \sigma_{2i}\sigma_{2j} )
+J_4\sigma_{1i}\sigma_{1j}\sigma_{2i}\sigma_{2j}
\right] , 
\label{AshTel}
\end{eqnarray}
can be described by the Dirac model with the interaction (\ref{Int}),
and with the random mass as well, 
if bond-randomness is included.\cite{DotDotAT}
As in the previous case, the interaction breaks the symmetry 
up to $n$ copies of O(2).
In passing, it should be noted that if the random bond variables $J_{ij}$ 
for spin 1 and spin 2 are independent, which is the case Dotsenko and Dotsenko 
mainly studied,\cite{DotDotAT}
critical behavior is slightly different form the one below.

Let us set $\gtilde_+=-\gtilde_-=\gtilde_M$ 
($\gtilde_V=0$) and $\gtilde_A=0$. Then, 
the scaling equations for this specific model are given by
\begin{eqnarray}
&&
\frac{d\gtilde_M}{dl}=-\gtilde_M^2+\gtilde\gtilde_M,
\nonumber\\
&&
\frac{d\gtilde}{dl}=-\gtilde\gtilde_M .
\label{RanMasRGE}
\end{eqnarray}
In the following subsections, we study the $\gtilde=0$ case
and $\gtilde\ne0$ case separately.

\subsection{Non-interacting case}

This subsection is devoted to the case $\gtilde=0$, which corresponds to 
the random bond Ising model.
As a solution of Eq. (\ref{RanMasRGE}), we have
\begin{eqnarray}
\gtilde_M=\frac{\gtilde_{M0}}{1+\gtilde_{M0}(l-l_0)},
\end{eqnarray}
where $\gtilde_{M0}$ is the initial coupling constant at $l=l_0$.
Substituting this into Eqs. (\ref{AnoDim}) and (\ref{EneSca}), 
we have
\begin{eqnarray}
\ln\frac{\Lambda}{E}=&&l-l_0+\frac{1}{2}
\ln\left[ 1+\gtilde_{M0}\left(l-l_0\right) \right] 
\nonumber\\
&&
\sim l-l_0+\frac{1}{2}\ln\left(l-l_0\right) ,
\end{eqnarray}
and together with (\ref{DosSca}), we obtain the following DOS 
\begin{eqnarray}
\rho(E)\sim\frac{E}{\Lambda}\ln\left(\frac{\Lambda}{E}\right) .
\label{RanMasDosFre}
\end{eqnarray}
The nonlinear sigma model for the class D has predicted\cite{SenFis,BSZ}
$|\ln E|$ behavior of DOS. 
Actually, on intermediate energy scales,
the $|\ln E|$ factor dominates the DOS in Eq. (\ref{RanMasDosFre}).

\subsection{Interacting case}

If the interaction is finite, there are two fixed points depending on the 
sign of the interaction $g_0$ as follows:
\begin{enumerate}
\item 
$\gtilde_M(l)\rightarrow0$ and $\gtilde(l)\rightarrow0$ 
for $\gtilde_0>0$.
\item $\gtilde_M(l)\rightarrow0$ and $\gtilde(l)\rightarrow \gtilde^*(<0)$ 
for $\gtilde_0<0$.
\end{enumerate}
This indicates that the Ising fixed point or, in other words, 
the class D is stable against the positive interaction.
In the case of the negative interaction, however,
it flows to fixed points depending on $g_0$, away from the class D.

The scaling equations in the case $\gtilde\ne0$ have been solved by 
Dotsenko and Dotsenko:\cite{DotDotAT} 
It may be convenient to derive first the relationship between 
$\gtilde_M$ and $\gtilde$: From (\ref{RanMasRGE})
\begin{eqnarray}
\gtilde_M(l)=\gtilde(l)\ln\left(\frac{\gtilde^*}{\gtilde(l)}\right) ,
\label{RanMasRel}
\end{eqnarray}
where 
\begin{eqnarray}
\gtilde^*=\gtilde_0\exp\left(\frac{\gtilde_{M0}}{\gtilde_0}\right)
=\frac{2g_0}{\pi}\exp\left(\frac{g_{M0}}{2g_0}\right) .
\label{RanMasDefGst}
\end{eqnarray}
This constant is a fixed point value of $\gtilde$ for negative $\gtilde_0$, 
as we shall see momentarily. 
These equations lead to
\begin{eqnarray}
\li\left(\frac{\gtilde^*}{\gtilde(l)}\right)=\gtilde^*(l-l_0)+
\li\left(\frac{\gtilde^*}{\gtilde_0}\right) ,
\end{eqnarray}
where $\li(x)$ is the logarithmic integral function.
In what follows, we solve the equations separately 
for positive and negative cases.

\subsubsection{The $g_0>0$ case}

As mentioned above,  $\gtilde^*/\gtilde(l)\rightarrow+\infty$ in this case.
By the use of the asymptotic expansion of the logarithmic integral function
\begin{eqnarray}
\li x\sim x\left( \frac{1}{\ln x} + \frac{1}{\ln^2x}+\dots \right) ,
\quad \mbox{for}\quad x\rightarrow+\infty ,
\end{eqnarray}
we can obtain the following asymptotic expression of 
the coupling constants
\begin{eqnarray}
&&
\gtilde(l)\sim\frac{g^*}{x\ln x}
\left( 1-\frac{\ln\ln x}{\ln x}+\frac{1}{\ln x} \right) ,
\nonumber\\
&&
\gtilde_M(l)\sim\frac{g^*}{x}\left( 1+\frac{1}{\ln x} \right),
\label{RanMasPosSol}
\end{eqnarray}
with $x=g^*l$. 
Eq. (\ref{EneSca}) is computed asymptotically as
\begin{eqnarray}
\ln\frac{\Lambda}{E}
\sim l+\frac{1}{2}\ln l+\ln\ln l +O\left(\frac{1}{\ln l}\right) .
\end{eqnarray}  
This equation, together with Eq. (\ref{DosSca}), leads to the following DOS,
\begin{eqnarray}
\rho(E)\sim \frac{E}{\Lambda}\ln\frac{\Lambda}{E}
\left( \ln\ln\frac{\Lambda}{E} \right)^2 .
\end{eqnarray}
As compared with the previous result (\ref{RanMasDosFre}), extra 
$\ln|\ln E|$ factor shows up due to 
the flow effect of the coupling constant $g$.  
It should be noted, however, that this fixed point belongs to the class D.

\subsubsection{The $g_0<0$ case}

The result obtained so far tells that universality class D is stable against 
the positive interaction. However, for the negative interaction, 
$\gtilde(l)$  flows to nonzero $\gtilde^*(<0)$ 
and a new universality class appears.
Note that $0<\gtilde^*/\gtilde(l)<1$, then we use the following formula
for the logarithmic integral function,
\begin{eqnarray}
\li x\sim \gamma+\ln\left| x-1\right| +\frac{x-1}{2}+\dots,
\quad \mbox{for}\quad  x\sim1 ,
\end{eqnarray} 
where $\gamma$ is Euler's constant.
Now we have
\begin{eqnarray}
&&
\gtilde\sim\frac{\gtilde^*}{1-e^x},
\nonumber\\
&&
\gtilde_M\sim\frac{\gtilde^*}{1-e^x}\ln(1-e^x),
\label{RanMasNegSol}
\end{eqnarray}
where $x$ is the same definition as the one in Eq. (\ref{RanMasPosSol}).
Therefore,  
Eq. (\ref{EneSca}) is computed as
\begin{eqnarray}
&&
\ln \frac{\Lambda}{E}\sim z l +O(e^{-|\gtilde^*|l}) , 
\nonumber\\
&&
z=1+\frac{\gtilde^*}{2}
=1-\frac{1}{\pi}|g_0|\exp\left( -\frac{g_{M0}}{|g_0|} \right)   , 
\end{eqnarray}  
and we reach
\begin{eqnarray}
\rho(E)\sim \left(\frac{E}{\Lambda}\right)^{(2-z)/z}.
\end{eqnarray}
The power-law behavior of the DOS tells us that 
that the criticality is controlled by the interaction, and 
the fixed point does not belong to the class D.

\subsection{Spin correlation functions}

So far we have calculated the scaling dimension of 
the ensemble-averaged two-point
correlation function of the fermion fields, 
and DOS as a result. 
We will calculate in this subsection spin correlation functions of 
the disordered Ashkin-Teller model.

With respect to the Majorana fermions $\chi$ defined as
\begin{eqnarray}
&&
\psibar_L=\frac{1}{\sqrt{2}}\left(\chi_{1L}-i\chi_{2L}\right), \quad
\nonumber\\
&&
\psi_L=\frac{1}{\sqrt{2}}\left(\chi_{1L}+i\chi_{2L}\right), 
\label{DefMaj}
\end{eqnarray}
and similar for R-moving fermion,  the random mass term
and the interaction term can be written as
\begin{eqnarray}
&&
\calH_M=iM\left( \chi_{1R}\chi_{1L}+\chi_{2R}\chi_{2L} \right) ,
\nonumber\\
&&
\calH_{int}=2g \chi_{1R}\chi_{1L}\chi_{2R}\chi_{2L} .
\label{RanMasMaj} 
\end{eqnarray}
The former equation tells us that the energy operator $\varepsilon(z,\zbar)$,
which couples with the mass, is 
$\varepsilon(z,\zbar)=\varepsilon_1(z,\zbar)+\varepsilon_2(z,\zbar)$ with 
$\varepsilon_j(z,\zbar)=i\chi_{jR}(z)\chi_{jL}(\zbar)$.
The operator product expansion of the energy operator
and the spin field $\sigma_{j}(z,\zbar)$, 
which is a scaling field of $\sigma_{ji}$
in Eq. (\ref{AshTel}), is\cite{Ber,KadBro}
\begin{eqnarray}
i\chi_{jR}\chi_{jL}(z,\zbar)\sigma_{j'}(w,\wbar)
=\frac{1}{4\pi}\frac{\delta_{jj'}}{|z-w|}\sigma_j(w,\wbar) .
\label{MajSpiOPE}
\end{eqnarray}
Note our normalization of the coordinate $z$ in Eq. (\ref{TwoPoiFun})
followed by $2\pi$. 

To calculate $\overline{\langle \sigma_1(L)\sigma_1(0)\rangle^N}$ and
$\overline{\langle \sigma_1\sigma_2(L)\sigma_1\sigma_2(0)\rangle^N}$,
let us consider the following two kinds of operators
\begin{eqnarray}
&&
\calO_1^{(N)}(z,\zbar)=\prod_{\alpha=1}^N\sigma_{1\alpha}(z,\zbar),
\nonumber\\
&& 
\calO_{12}^{(N)}(z,\zbar)
=\prod_{\alpha=1}^N\sigma_{1\alpha}(z,\zbar)\sigma_{2\alpha}(z,\zbar) .
\label{RanMasSpi}
\end{eqnarray}
We assume that there are enough number of replicas $n>N$ and 
compute the OPE between (\ref{RanMasSpi}) and four fermi operators
in Eq. (\ref{RepHam}) 
using the basic OPE between Majorana fermions and
the spin field (\ref{MajSpiOPE}). 
It turns out that the results depend on $N$ only,
and we obtain the following anomalous dimensions
of the fields $\calO_1$ and $\calO_{12}$, respectively,
\begin{eqnarray}
&&
\gamma_1^{(N)}=\frac{N}{8}-\frac{N(N-1)}{16}\gtilde_M,
\nonumber\\
&&
\gamma_{12}^{(N)}=\frac{2N}{8}-\frac{2N(2N-1)}{16}\gtilde_M-\frac{N}{8}\gtilde .
\end{eqnarray}
It should be noted that $\gamma_1^{(N)}$ is completely the same as the one for 
the random-bond Ising model.\cite{Ber}
Using Eqs. (\ref{RanMasPosSol}) and (\ref{RanMasNegSol}), we have
the correlation function 
$\overline{\langle \sigma_1(L)\sigma_1(0)\rangle^N}$
at large distances,
\begin{eqnarray}
&&
\overline{\langle \sigma_1(L)\sigma_1(0)\rangle^N}
\nonumber\\&&  
=
\left\{
\begin{array}{lr} 
\displaystyle{
\left(\frac{L_0}{L}\right)^{ N/4 } 
\left( \ln \frac{L}{L_0} \right)^{ N(N-1)/8 } 
          }& (g_0>0)
\\
\displaystyle{
\left(\frac{L_0}{L}\right)^{ N/4 } 
           }& (g_0<0) .
\end{array}
\right. 
\end{eqnarray}
As expected from the analysis in the last subsection,
the correlation function of this type is not affected by the positive 
interaction. In the case of the negative one, however,
the logarithmic correction do not remain and correlation 
function is just the one for the pure Ising model.
Contrary to $\gamma_1^{(N)}$, since $\gamma_{12}^{(N)}$ depends on
$\gtilde$ explicitly, it follows that the interaction effects are more manifest.
Actually, we obtain 
\begin{eqnarray}
&&
\overline{\langle \sigma_1\sigma_2(L)\sigma_1\sigma_2(0)\rangle^N}
\nonumber\\&& \nonumber\\&& 
=
\left\{
\begin{array}{lr}
\displaystyle{
\left(\frac{L_0}{L}\right)^{ N/2 } 
\left( \ln \frac{L}{L_0} \right)^{ N(2N-1)/4} 
              }
&\\ \hspace{30mm}\times 
\displaystyle{
\left( \ln\ln \frac{L}{L_0} \right)^{ N/4 }
            } & (g_0>0)
\\ & \\
\displaystyle{
\left(\frac{L_0}{L}\right)^{ N \left(1-\gtilde^*/2 \right)/2 } 
          }& (g_0<0) .
\end{array}
\right. 
\nonumber\\&& 
\end{eqnarray}
Namely, due to the interaction, 
extra $\ln\ln L$ factor shows up in the former case, 
and the exponent depends manifestly on the 
coupling constant of the interaction 
through the relation (\ref{RanMasDefGst}) in the latter case.

\section{The random scalar potential}

Setting $g_+=g_-=g_V$ as well as $g_A=0$ we have
\begin{eqnarray}
&&
\frac{d\gtilde_V}{dl}=\gtilde_V^2+\gtilde\gtilde_V,
\nonumber\\
&&
\frac{d\gtilde}{dl}=\gtilde\gtilde_V .
\label{RanScaRGE}
\end{eqnarray}
As discussed by Ludwig et.al.,\cite{LFSG} 
the case with random scalar potential only 
($\gtilde=0$) belong to the class AII 
because of time-reversal symmetry.\cite{AltZir}
It follows that the model is expected to flow to the fixed point
of the symplectic nonlinear sigma model,  
although in the present weak-coupling approach, 
$\gtilde_V$ is marginally relevant and
flows to a  strong coupling regime. 
With finite $\gtilde_0>0$, Eq. (\ref{RanScaRGE}) tells us that 
$\gtilde_V$ and $\gtilde$ also flows to infinity.
In the case $\gtilde_0<0$, however, the model has a fixed point
$\gtilde_V\rightarrow0$ and $\gtilde\rightarrow g^*$, where $\gtilde^*$ is 
specified momentarily. 

In this section, we examine only the negative $g_0$ case,
and follow a similar calculation to the previous section 
of the random mass model.
First of all, we note the relationship
\begin{eqnarray}
&&
\gtilde_V(l)=-\gtilde\ln
\left( \frac{\gtilde^*}{\gtilde(l)} \right), 
\nonumber\\
&&
\gtilde^*=\gtilde_0\exp \left( -\frac{\gtilde_{V0}}{\gtilde_0} \right),
\end{eqnarray}
which correspond to Eqs. (\ref{RanMasRel}) and (\ref{RanMasDefGst}).
Therefore, the asymptotic solutions for large $l$ are given by
\begin{eqnarray}
&&
\gtilde=\frac{\gtilde^*}{1+e^{\gtilde^*l}},
\nonumber\\
&&
\gtilde_V=-\frac{\gtilde^*}{1+e^{\gtilde^*l}}\ln(1+e^{\gtilde^*l}).
\end{eqnarray}
Using these, we finally end up with
\begin{eqnarray}
\rho(E)\sim \left(\frac{E}{\Lambda}\right)^{(2-z)/z} ,
\end{eqnarray}
where
\begin{eqnarray}
z=1+\frac{\gtilde^*}{2}
=1-\frac{|g_0|}{\pi}\exp\left( \frac{ g_{V0} }{ 2|g_0| } \right) .
\end{eqnarray}
It turns out that negative interaction yields 
nontrivial fixed point dependent on
the strength of the interaction,
which does not belong to AII.

\section{The random potentials with chiral symmetry}

So far we have studied some fixed points of the RG equations (\ref{RGeqs}).
If we allow non-hermitian Hamiltonian, 
there is still another fixed point with $g_+=0$.   
Guruswamy et. al. \cite{GLL} have proposed a model 
with a random imaginary vector potential,  a random imaginary scalar potential 
and a random real mass term, 
as a fermion model equivalent to 
the random phase sine-Gordon model.\cite{Ber,CarOst}
In this section we study how interaction affects the critical behavior.
Let us add the interaction (\ref{Int}) to the Hamiltonian 
which Guruswamy et. al. have studied
\begin{eqnarray}
\calH=\psibar i\gamma_\mu\left(\partial_\mu+A_\mu\right)\psi
+M\psibar\sigma_3\psi+iV\psibar\psi+\calH_{int} .
\label{ChiHamOne}
\end{eqnarray}
This is basically the same model as the random phase sine-Gordon model
with arbitrary boson coupling $K$.\cite{Ber} 
The RG equations of this model are given by setting $g_A\rightarrow-g_A$,
$g_+=0~(g_V=-g_M<0)$, and $g_-=-2g_M\rightarrow -g_-$,
\begin{eqnarray}
&&
\frac{d\gtilde_A}{dl}=\gtilde_-^2,
\nonumber\\
&&
\frac{d\gtilde_-}{dl}= \gtilde\gtilde_- ,
\nonumber\\
&&
\frac{d\gtilde}{dl}= -\gtilde\gtilde_- .
\label{ChiRGE}
\end{eqnarray}

This model may not be suitable for the computation of the DOS
because of its non-hermiticity. 
A simple extension, however, to 
a two Dirac fermion model yields a hermitian Hamiltonian 
with the same RG equations in the following way;
\begin{eqnarray}
\calH_{12}=\calH_1+\calH_2 ,
\end{eqnarray}
where $\calH_j$ is the Hamiltonian (\ref{ChiHamOne}) for the fermion 
$\psi_j$ and $\psibar_j$.
Next, make the transformation,
\begin{eqnarray}
\begin{array}{ll}
\psibar_{1R}\rightarrow\psibar_{1R}, & \psi_{1R}\rightarrow\psi_{1R},\\
\psibar_{1L}\rightarrow\psibar_{2L},  & \psi_{1L}\rightarrow\psi_{2L},\\
\psibar_{2R}\rightarrow\psi_{2R},     & \psi_{2R}\rightarrow\psibar_{2R},\\
\psibar_{2L}\rightarrow\psi_{1L},      & \psi_{2L}\rightarrow\psibar_{1L},
\end{array}
\end{eqnarray}
which yields no change to the partition function up to some constants, 
and, at the same time, the transformation as well
\begin{eqnarray}
A_1\rightarrow A_2, \quad A_2\rightarrow -A_1 ,
\end{eqnarray}
which keeps the probability distribution (\ref{ProDis}) invariant. 
Then, these transformation yields
\begin{eqnarray}
\calH_{12}\rightarrow&&
 \psibar_1i\gamma_\mu\left(\partial_\mu-iA_\mu\right)\psi_1
+\psibar_2i\gamma_\mu\left(\partial_\mu+iA_\mu\right)\psi_2
\nonumber\\&& 
+B\psibar_1\sigma_3\psi_2+\bar{B}\psibar_2\sigma_3\psi_1
\nonumber\\&& 
+2g\left(J_{1R}J_{2L}+J_{2R}J_{1L}\right) ,
\label{ConHatHam}
\end{eqnarray}
where $B$ is defined here as $B=M+iV$
and $J_{jR(L)}=\psibar_{jR(L)}\psi_{jR(L)}$
denotes the current of the $j$th fermion.
This Hamiltonian is actually hermitian, and
probability distribution of $A_\mu$ is the same as in (\ref{ProDis}), and
$\overline{B(x)\bar{B}(y)}=g_-\delta(x-y)$. 

Without the interaction $(g=0)$ in Eq. (\ref{ConHatHam}), 
this Hamiltonian is just the continuum limit of 
the random hopping fermion model on a square lattice with $\pi$-flux 
per plaquette,\cite{Hat}
or the one on a honeycomb lattice.
It should be noted that the interaction term in the Dirac 
Hamiltonian (\ref{ConHatHam}) is not due to interactions, 
but due to e.g., annealed randomness, of the original 
lattice models.\cite{Kir}
Actually, annealed randomness for
quenched random systems have turned out to play an important role 
in metal-insulator transitions.\cite{Kir}
With $n$ replicas, this extended model 
belongs to the class BDI when $g=0$, and has a global U($2n$) symmetry, 
which include chiral  U($n)\times$U($n$) symmetry.
To see this, it may be convenient to write the above Hamiltonian
with $g=0$ as follows;
\begin{eqnarray}
\calH_{12}=&&
 \Psibar_R2(\partial_\zbar + A_\zbar) \Psi_R
+\Psibar_L2(\partial_z + A_z) \Psi_L
\nonumber\\&& 
-iB\Psibar_L\Psi_R+i\bar B\Psibar_R\Psi_L ,
\end{eqnarray}
where $\Psibar_R=(\psibar_{1R},\psibar_{2R})$, 
$\Psi_R^t=(\psi_{1R},\psi_{2R})$, and similar for $\Psibar_L$ and $\Psi_L$. 
It is readily seen that the replicated model is invariant 
under the transformation 
\begin{eqnarray}
&&
\Psibar_R\rightarrow \Psibar_RU^\dagger, \quad\Psi_R\rightarrow U\Psi_R,
\nonumber\\
&&
\Psibar_L\rightarrow \Psibar_LU^\dagger, \quad\Psi_L\rightarrow U\Psi_L,
\label{BDIsym}
\end{eqnarray}
with
$2n\times 2n$ unitary matrix $U$.
Of course, the interaction breaks this symmetry, as in the previous cases.
Interestingly, this Hamiltonian has another symmetry.
\begin{eqnarray}
\begin{array}{ll}
\Psibar_R\rightarrow \Psibar_Re^{-i\theta} ,\quad 
&
\Psi_R\rightarrow e^{i\theta}\Psi_R,
\\
\Psibar_L\rightarrow \Psibar_L e^{i\theta}, 
&
\Psi_L\rightarrow e^{-i\theta}\Psi_L,
\end{array}
\end{eqnarray} 
with 
\begin{eqnarray}
\bar B\rightarrow \bar Be^{2i\theta},\quad B\rightarrow e^{-2i\theta} B.
\end{eqnarray}
This symmetry as well as (\ref{BDIsym}) allows us to  determine 
the $g_A$-dependence of the correlation functions.\cite{Ber,GLL}

We study this model 
perturbed by $-z(\psibar_1\psi_1+\psibar_2\psi_2)$ to derive 
the scaling function of the DOS.
The anomalous dimension of this operator is
\begin{eqnarray}
\gamma=1+\frac{\gtilde_A(l)}{4}+\frac{\gtilde_-(l)}{2} ,
\end{eqnarray}
and the scaling equations are the same as Eqs. (\ref{ChiRGE}).

\subsection{Non-interacting case}
To begin with, we review the results known so far 
for $\gtilde_0=0$ case.\cite{Ber,GLL}
The scaling equations for $\gtilde_A$ and $\gtilde_-$ in (\ref{ChiRGE}) 
are easy to integrate, giving the flow of the coupling constants,
\begin{eqnarray}
\gtilde_B(l)=\gtilde_{-0}, \quad 
\gtilde_A(l)=\gtilde_{A0}+\gtilde_{-0}^2(l-l_0),
\label{ChiNonIntSol}
\end{eqnarray}
where $\gtilde_{A0}$ and $\gtilde_{-0}$ are the initial coupling constants.
Substituting theses solutions to Eq. (\ref{EneSca}), we obtain
\begin{eqnarray}
\ln\frac{\Lambda}{E}=z(l-l_0)+\frac{y}{2}(l-l_0)^2,
\end{eqnarray}
where 
\begin{eqnarray}
&&
z=1+\frac{\gtilde_{A0} }{4}+\frac{\gtilde_{-0}}{2}=
1+\frac{ g_{A0} +g_{M0} }{\pi} ,
\nonumber\\
&&
y=\frac{\gtilde_{-0}^2}{4}=\frac{g_{M0}^2}{\pi^2}.
\label{DefZY}
\end{eqnarray}
Combining this equation with Eq. (\ref{DosSca}), 
we reach the famous expression 
of the DOS for the class BDI, which Gade derived 
for the first time,\cite{GadWeg}
\begin{eqnarray}
\rho(E)\sim\frac{\Lambda}{E}\exp
\left[
-\frac{2z}{y}\left(\sqrt{1+\frac{2y}{z^2} \ln\frac{\Lambda}{E}  }-1\right)
\right] .
\label{ChiDos}
\end{eqnarray}
This formula shows that the DOS diverges towards the zero energy.
However, the crossover energy scale under which the singularity of the
DOS emerges is estimated as  
\begin{eqnarray}
E_{cr}\sim\Lambda e^{-z^2/(2y) }
\sim \Lambda\exp \left( -\frac{\pi^2}{2g_{M0}^2} \right) ,
\end{eqnarray} 
or in terms of the length scale, the crossover length is given by
$L_{cr}\sim L_0\exp\left( 2\pi^2/g_{M0}^2 \right) $,
followed from Eq. (\ref{ChiNonIntSol}).
Namely, the crossover energy (length)
is exponentially small (large) for weak randomness.
Considering this, we expect
\begin{eqnarray}
\rho(E)\sim
\left\{
\begin{array}{lr}
\displaystyle{
\frac{\Lambda}{E}
\exp\left( -\sqrt{\frac{8}{y} \ln\frac{\Lambda}{E} }\right)  
                 } &\quad  (E<<E_{cr})
\\
\displaystyle{
\left( \frac{E}{\Lambda}\right)^{(2-z)/z} 
                 } & (E>>E_{cr}) .
\end{array}
\right.
\end{eqnarray}
Previous numerical studies \cite{Hat} observed a power-law behavior 
rather than the singular divergence even quite near the zero energy. 
In the regime of their study,
$E_{cr}$ may be  too small to yield
the divergent DOS,  or 
the system size is much smaller than the crossover length.
Actually,  Ryu and Hatsugai have recently been successful in observing 
the divergent DOS by studying 
quite large systems.\cite{RyuHat}
The DOS calculated by them may be described directly by Eq. (\ref{ChiDos}).

\subsection{Interacting case}

Next consider the case with finite interaction $g_0\ne0$.
First, it should be noted that $\gtilde_-+\gtilde$ is conserved 
under the scale transformation,
\begin{eqnarray}
\gtilde_-+\gtilde=\gtilde_{-0}+\gtilde_0\equiv c .
\end{eqnarray}
Therefore, only two equations among three in Eqs. (\ref{ChiRGE}) are 
essentially independent.
By using $c$, the equation for $\gtilde_-$ can be converted into
\begin{eqnarray}
\frac{d\gtilde_-}{dl}=-\gtilde_-(\gtilde_--c) .
\end{eqnarray}
It is easy to solve this equation, 
\begin{eqnarray}
\gtilde_-(l)=\frac{c}{ 1+(\gtilde_0/\gtilde_{-0})e^{-c(l-l_0)} } ,
\label{GmCnonz}
\end{eqnarray}
for nonzero $c$, whereas for $c=0$, 
\begin{eqnarray}
\gtilde_-(l)=\frac{\gtilde_{-0}}{1+\gtilde_{-0}(l-l_0)} .
\label{GmCz}
\end{eqnarray}

Directly from these solutions or 
by analyzing the beta functions in Eq. (\ref{ChiRGE}),
it is not difficult to observe that there are three kinds of fixed points.
\begin{enumerate}
\item 
$\gtilde_-(l)\rightarrow c~(\equiv\gtilde_{-0}+\gtilde_0)$ and 
$\gtilde(l) \rightarrow0$ for $c>0$.
\item 
$\gtilde_-(l)\rightarrow 0$ and 
$\gtilde(l) \rightarrow0$ for $c=0$.
\item
$\gtilde_-(l)\rightarrow 0$ and 
$\gtilde(l) \rightarrow c$ for $c<0$.
\end{enumerate}
In the following subsections, we study the cases separately.

\subsubsection{The case $c>0$}

As $\gtilde(l)\rightarrow 0$, this fixed point is just the same as that of
the non-interacting case.
The flow of $\gtilde_A(l)$ is easy to integrate in Eq. (\ref{ChiRGE}) by the
use of Eq. (\ref{GmCnonz}), and
for large $l$ the coupling constants are approximately given by 
\begin{eqnarray}
&&
\gtilde_-(l)\sim  c,
\nonumber\\
&&
\gtilde_A(l)\sim c^2(l-l_0)+\gtilde_{A0}
-c\ln\left( \frac{\gtilde_0}{\gtilde_{-0}}+1\right)-\gtilde_0 .
\label{ChiIntSol}
\end{eqnarray}
Substituting these into Eq. (\ref{EneSca}), we have
\begin{eqnarray}
\ln\frac{\Lambda}{E}\sim z'(l-l_0)+\frac{y'}{2}(l-l_0)^2 ,
\end{eqnarray}
with
\begin{eqnarray}
&&
z'=z+\frac{1}{2\pi} ,
\left[
g_0-\left(g_{M0}+g_0\right)\ln \left( \frac{g_0}{g_{M0} } +1\right)
\right] ,
\nonumber\\
&&
y'=c^2/4=\left( \frac{g_0+g_{M0}}{\pi} \right)^2 ,
\end{eqnarray}
where $z$ is defined in Eq. (\ref{DefZY}). Above equation 
leads to Eq. (\ref{ChiDos}) with modified parameters $y'$ and $z'$.

\subsubsection{The case $c=0$}

We expect, at first sight, that this case also flows to 
the same fixed point as above,
since $\gtilde(l)\rightarrow0$. However, the integration of $\gtilde_A$ gives
\begin{eqnarray}
\gtilde_A(l)=-\frac{\gtilde_{-0}}{1+\gtilde_{-0}(l-l_0)}
+\gtilde_{-0}+\gtilde_{A0} .
\end{eqnarray}
In the previous cases, the $l$ term of $\gtilde_A$ 
in Eqs. (\ref{ChiNonIntSol}) and (\ref{ChiIntSol}) is responsible 
for the divergent DOS at the band center, which is missing in the present case.
Now we have
\begin{eqnarray}
\ln\frac{\Lambda}{E}\sim z(l-l_0)+\frac{1}{4}\ln(l-l_0) ,
\end{eqnarray}
with
\begin{eqnarray}
z=1+\frac{\gtilde_{A0}+\gtilde_{-0}}{4} 
=1+\frac{ 2g_{A0}+g_{M0} }{2\pi} .
\end{eqnarray}
This leads to
\begin{eqnarray}
\rho(E)\sim
\left(\frac{E}{\Lambda}\right)^{(2-z)/z}
\left( \ln\frac{\Lambda}{E} \right)^{1/2z} .
\end{eqnarray}
The DOS in this case shows a power-law behavior with a logarithmic correction.
Although this fixed point should be in the class BDI, the behavior of the DOS 
is quite different.

\subsubsection{The case $c<0$}

As we have seen in the last two subsections, the class BDI is stable 
against the interaction in the case $c\ge0$, although DOS is different
from that of BDI for $c=0$. Contrary to this, the present case leads to
a different fixed point which does not belong to BDI any longer.
Asymptotically for large $l$ 
the flows of the coupling constants are computed as
\begin{eqnarray}
&&
\gtilde_-(l)\sim  0,
\nonumber\\
&&
\gtilde_A(l)\sim \gtilde_{A0}+\gtilde_0
-c\ln\left|1+ \frac{\gtilde_0}{\gtilde_{-0}} \right| +O(e^{-|c|l}) .
\end{eqnarray}
Therefore we obtain
\begin{eqnarray}
\ln\frac{\Lambda}{E}\sim z(l-l_0) ,
\end{eqnarray}
where
\begin{eqnarray}
z=&&1+\frac{\gtilde_{A0}+\gtilde_0}{4}
-\frac{1}{4}c\ln\left|1+ \frac{\gtilde_0}{\gtilde_{-0}} \right| 
\nonumber\\
=&&1+\frac{2g_{A0}+g_{M0}}{2\pi}
+\frac{|g_{M0}+g_0|}{2\pi}\ln\left|1+ \frac{g_0}{g_{M 0}} \right|  .
\end{eqnarray}
This equation leads to 
\begin{eqnarray}
\rho(E)\sim\left( \frac{E}{\Lambda} \right)^{(2-z)/z} .
\end{eqnarray}

\section{Summary and discussions}

In this paper, we have investigated the interaction effects on the
disordered Dirac fermion model separately with a random vector potential,
a random mass, a random scalar potential, and their special combination
with chiral symmetry. 
They belong to the classes 
AIII, D, AII and BDI, respectively, when the interaction is zero.
It turns out that some fixed points are stable against the interaction
in the sense that the coupling constant of the interaction flows to zero
in the RG, whereas others are unstable and flow
to fixed points which do not belong to those classes listed-above.
Thus, it turns out that interactions play an important role 
in the symmetry properties of localization problems.

So far we have studied several specific fixed points. In what follows,
we would like to discuss briefly the case where all kinds of disorder are 
simultaneously present.
As was shown by Ludwig et. al., \cite{LFSG}
the fixed line of the random vector 
potential as well as the fixed point of a random mass are unstable and flow
to unknown strong-coupling regime on generic initial conditions. 
In the present case with the positive interaction,  
the situation is quite similar to  
the case of non-interacting case.
Numerical calculation of the scaling equations (\ref{RGeqs})
shows that the coupling constants flow to infinity.
The negative interaction, however, can give rise to a different situation.
Below a certain critical $g=g_c<0$, 
which may depend on the other disorder strength, 
the solution of the scaling equations flow to a fixed point, 
which is quite similar to the fixed line of the model with 
the random vector potential and the interaction only, discussed in Sec. III.
Namely, after the flow, 
$g_A$ and $g$ become finite constants whereas $g_\pm\rightarrow0$.
Therefore, we could claim that the fixed line formed 
by the random vector potential and the marginal interaction 
is one of generic critical points for disordered Dirac fermion model
with randomness and interaction. 
Other fixed points are in a strong coupling 
regime and are not accessible by a weak-coupling approach.

As it has turned out that an interaction plays a role 
in a simple disordered model like Dirac fermions examined in this paper, 
it is quite interesting to study an interplay between interactions 
and randomness for more complicated 
systems, especially those belonging to the new universality classes.

\acknowledgements
The author would like to thank J. T. Chalker for valuable discussions 
in the early stage of the work. 
He is supported in part by the Japan Society for 
the Promotion of Science, and the Yamada Science Foundation.


\end{multicols}

\end{document}